# Personalized and Constructive Feedback for Computer Science Students Using the Large Language Model (LLM)

Javed Ali Khan[1,*], Muhammad Yaqoob[1], Mamoona Tasadduq[2], Hafsa Shareef Dar[3], Aitezaz Ahsan[1]
[1]Department of Computer Science, University of Hertfordshire, Hatfield, UK
[2]Information Technology University, Lahore, Pakistan
[3]Department of Software Engineering, Faculty of Computing and IT, University of Gujrat, Pakistan

Corresponding author: j.a.khan@hert.ac.uk

**Abstract:** The evolving pedagogy paradigms are leading toward educational transformations. One fundamental aspect of effective learning is relevant, immediate, and constructive feedback to students. Providing constructive feedback to large cohorts in academia is an ongoing challenge. Therefore, academics are moving towards automated assessment to provide immediate feedback. However, current approaches are often limited in scope, offering simplistic responses that do not provide students with personalized feedback to guide them toward improvements. This paper addresses this limitation by investigating the performance of Large Language Models (LLMs) in processing students' assessments with predefined rubrics and marking criteria to generate personalized feedback for in-depth learning. We aim to leverage the power of existing LLMs for Marking Assessments, Tracking, and Evaluation (LLM-MATE) with personalized feedback to enhance students' learning. To evaluate the performance of LLM-MATE, we consider the Software Architecture (SA) module as a case study. The LLM_MATE approach can help module leaders overcome assessment challenges with large cohorts. Also, it helps students improve their learning by obtaining personalized feedback in a timely manner. Additionally, the proposed approach will facilitate the establishment of ground truth for automating the generation of students' assessment feedback using the ChatGPT API, thereby reducing the overhead associated with large cohort assessments.

## 1. Introduction:

Information technology has rapidly evolved over the past decade and affected many areas, including education (Alier et al., 2024). This emerges as a new concept in Smart Education, a learning and teaching paradigm that utilizes Artificial Intelligence (AI), the Internet of Things (IoT), and other technologies for enhanced teaching and learning experiences (Badshah et al., 2023). It provides potential learners, teachers, and other stakeholders with a digital platform for improved learners' motivation and engagement (Zhu et al., 2016). After the COVID-19 pandemic outbreak in 2020, the need for smart education has become more significant and essential. This leads many educational institutions to utilize smart education platforms to transfer teaching and learning activities online using AI and machine learning (ML) (Dimililer et al., 2023). Similarly, it allows educational researchers and practitioners to improve learners' engagement in teaching and learning by adopting active learning approaches using smart education platforms.

Considering this pressing need, educational researchers and vendors have started developing various educational AI-based applications supporting various teaching and learning activities by employing state-of-the-art machine AI approaches (Wang et al., 2024). Similarly, researchers applied AI to various educational activities, including administration, instruction, learning, and instructors' and students' performances (Chen et al., 2020). Some of the recent AI-based applications in education include Duolingo[1], which uses state-of-the-art AI technology to help learn and improve languages in a fun and interactive way. Similarly, Acorn[2] and Absorb[3] are examples of some of the best smart learning management systems (LMS) powered by AI that provide skill development, smart content creation and recommendations, learning analytics, administration, etc., features to support teaching and learning (Leh, 2022). More recently, researchers have started using LLMs to automate and improve various domains with greater success, including software engineering (Hou et al., 2023) and education (Wang et al., 2024) (Alier et al., 2024) activities. For example, Khan Academy developed Khanmigo[4], powered by LLM, i.e., ChatGPT, which offers personalized learning across various subjects, including programming, maths, and language learning. Also, Khan et al. recently employed ChatGPT API to effectively classify end-user feedback into various emotion types, helping software engineers better understand the end-user grudges with the software applications (Khan et al., 2024). The growing capabilities of LLMs in processing and analyzing textual and image data for valuable insight show great potential in automating various educational activities, particularly assessment evaluation and generating personalized feedback.

Previously, researchers have proposed various AI-enabled approaches to gauge the learning outcome of the learners. For example, Fu et al. investigated the usage of AI-enabled automatic scoring tools for language learners and their impact on learners' continuous learning. Similarly, Lu and Cutumisu used state-of-the-art deep learning algorithms to generate an automated essay score (Lu & Cutumisu, 2021). Also, Liang et al. proposed a learning activity-based approach to engineer features that may enable accurate at-risk predictions and meaningful feedback generation (Liang et al., 2024). Also, most automated assessment approaches in the literature are based on formative assessments to gauge the learners learning (González-Calatayud et al., 2021). However, current approaches are often limited in scope, offering simplistic responses without providing students with personalized feedback that guides them toward improvements. Moreover, it is evident in the literature that AI-based approaches for automated assessment can potentially improve student learning by providing constructive and timely feedback (González-Calatayud et al., 2021). Therefore, we aim to leverage the power of LLM (ChatGPT) to evaluate students' summative assessments using predefined rubrics to provide constructive personalized feedback and mark the assessment based on its quality and correctness. In particular, we are interested in answering the following research questions. **RQ1**: Can LLM (ChatGPT) assist teachers in providing personalized student assessments?; and **RQ2**: How accurate and reliable is the AI-generated personalized feedback for Computer Science students?

---

[1] https://spectrum.ieee.org/duolingo accessed on 15-November 2024
[2] https://acorn.works/ accessed on 15-November 2024
[3] https://talentedlearning.com/lms/absorb-lms/ accessed on 15-November 2024
[4] https://www.khanmigo.ai/ accessed on 10-November 2024

The structure of the paper is as follows: Section 2 elaborates on the related work, section 3 discusses the proposed methodology, and section 4 demonstrates the results and discussion. Section 5 concludes the paper and discusses the future directions.

## 2. Related Work

Recent research in automated assessment and feedback tools has aimed at bridging the gap between traditional methods and more sophisticated, data-driven approaches. Several studies have highlighted the potential of AI-driven techniques to provide more relevant feedback and personalized learning experiences.

Biswas et al. proposed a machine learning (ML)-based intelligent real-time feedback system (Biswas et al., 2023). The authors leverage the classification model in ML techniques to provide timely and personalized feedback within blended learning environments. The authors emphasize the system's potential to enhance the learning experience by delivering immediate insights to students and teachers. Gutierrez and Atkinson investigate the adaptive feedback method in the context of intelligent tutoring systems (Gutierrez & Atkinson, 2011). The authors proposed a novel three-stage model for providing feedback, which included error detection, feedback strategy selection based on the error, and automatic feedback generation using machine learning. The model aims to enhance learning outcomes by tailoring feedback strategies to different student errors and scenarios. Van der Merwe et al., integrated the feedback mechanism in LMS and evaluated the framework using quantitative and qualitative methods. The survey participants expressed a satisfaction level of 94% regarding the feedback content and 95% satisfaction concerning the confidential handling of their academic data. Similarly, Wecks et al. analyze the student's usage of ChatGPT on their exam performances. Their findings report that students using ChatGPT score comparatively lower. However, it helps in learning and engagement (Wecks et al., 2024).

Bimba et al. critically reviewed 20 research papers on adaptive feedback by identifying interesting insights and advantages of incorporating continuous feedback for improved learning (Bimba et al., 2017). The authors observed that students' characteristics and knowledge levels are critical for providing relevant feedback. They also found that latent semantic analysis (LSA) and parsers are common techniques for implementing adaptive feedback. In contrast, questionnaires, pre-tests, post-tests, and log data analysis are used to evaluate the model. The authors concluded that adaptive feedback seems to be more easily implemented in the programming domain due to the logical and procedural nature of the programming domain. Another comprehensive literature review by Maier et al. conducted 39 studies on personalized feedback in digital learning environments from the last decade, covering microscale to macroscale aspects (Maier et al., 2020). The authors observed that the adaptive feedback sources primarily revolve around present knowledge levels and data related to learning behaviors. Additionally, they found limited research attention in other interesting data sources such as emotional state assessments, progress metrics, learning objectives, and personality traits. The authors also highlighted the need to use AI for learner-specific feedback and integrate learner disposition data for improved customization.

These research findings emphasize the importance of incorporating sophisticated feedback mechanisms within educational systems, enhancing their functionality and effectiveness. By seamlessly integrating these advanced feedback tools with current platforms, educators are better equipped to provide comprehensive support and guidance to students, fostering a more dynamic and engaging learning environment. These research approaches mainly focus on formative assessments or are limited to certain domains, such as assessing language skills. In contrast, we aim to utilize the power of ChatGPT in assessing student summative submissions, generating constructive, detailed, and personalized feedback, and evaluating its efficacy for automating the process with domain experts.

## 3. Proposed Approach

Figure 1 depicts the proposed LLM-MATE approach, which comprises four steps. The approach is used to identify the efficacy of LLMs, i.e., ChatGPT, in evaluating learners' multi-model assessments using ChatGPT prompts for constructive and personalized feedback. Each methodological step is elaborated in detail below.

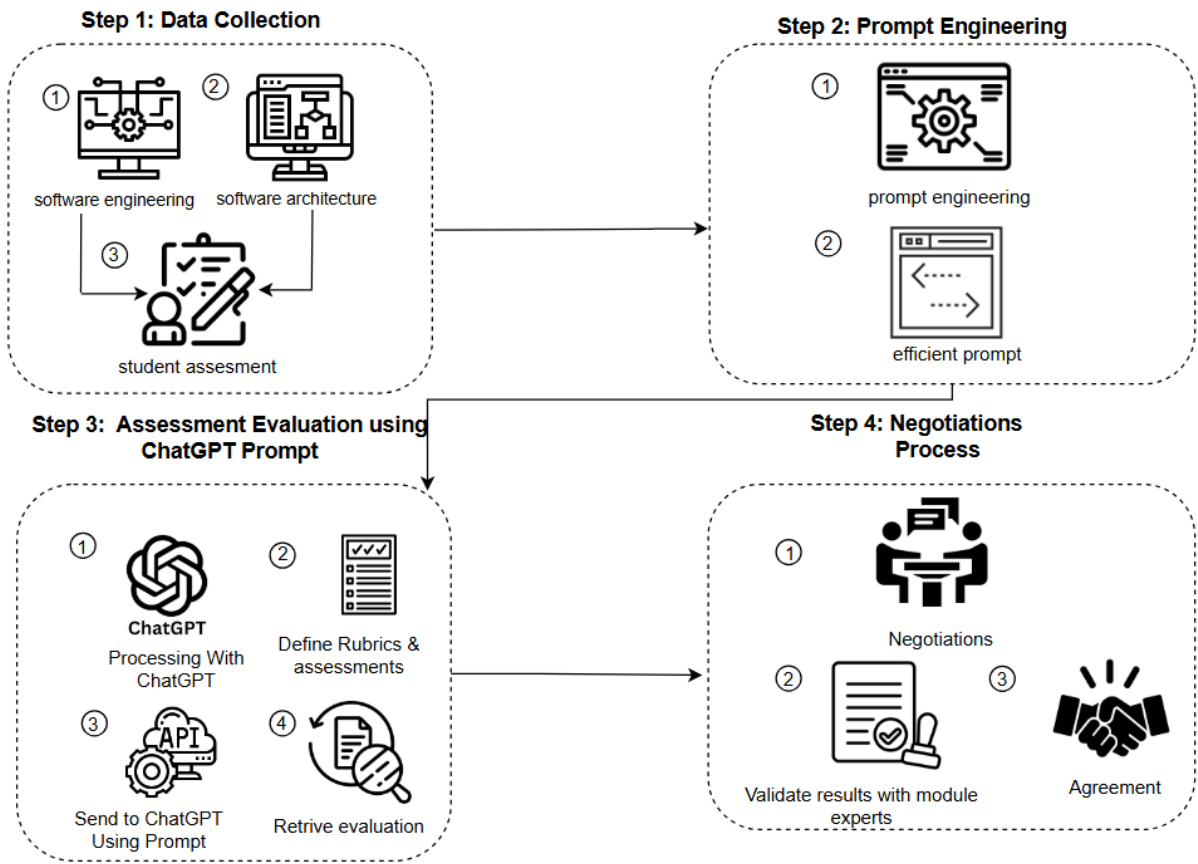

**Figure 1: Proposed ChatGPT prompt-based approach for assessment evaluation.**

**3.1 Data Collection:** To develop the proposed LLM-MATE approach, we gathered anonymized students' assessment data for the Software Architecture (SA) modules. The assessment landscape for the module covers essays, images, and programming assessments for delivering meaningful, constructive, and personalized feedback to the students. We selected the module as the student assessment involves both textual and figures, making it a better use case for identifying the performance of ChatGPT in generating constructive and personalized feedback by understanding and evaluating both text and figures. For the SA module, student assessments are designed, where students will submit different software engineering diagrams, such as use cases and class diagrams, with their corresponding rationale for the case study. We have obtained ethical approval to get the students' prior consent to use their assessments to be evaluated with the ChatGPT prompt. Moreover, the data collected from the students will be stored on the UH secure server for further processing.

**3.2 Prompt Engineering:** Before executing the proposed LLM-MATE approach, it is important to develop a customized prompt for generating effective end-user feedback to improve and enhance the student's learning experiences. Also, the effectiveness of ChatGPT response generation is highly dependent on well-structured prompts, which guide the ChatGPT in yielding relevant and accurate feedback (White et al., 2024). Prompt engineering will be employed to optimize responses and maximize relevance. Key strategies include:

**Domain Restriction**: Using well-structured prompts, the ChatGPT will be instructed to constrain analysis within specific parameters. This will ensure that responses remain relevant to the given domain, filtering out irrelevant information. This approach will improve the quality of feedback by keeping it focused on the content and context of student submissions.

**Personalized Feedback Generation:** Customized prompts will enable the ChatGPT to analyze each submission's strengths, areas for improvement, and recommendations for correction. This applies to text, images, and programming/algorithms assessments. The ChatGPT prompt will analyze student response patterns to deliver feedback highlighting aspects that require attention, such as conceptual understanding or syntactical clarity.

**Iterative Testing and Refinement:** Prompts will undergo extensive iterative testing to ensure consistent output quality across various assessment types. This testing will establish optimal prompt structures for generating constructive feedback. Through iterative refinement, prompts will be optimized to yield responses that balance specificity with clarity.

**Error Identification:** The ChatGPT prompt will be designed to identify errors in student submissions by pointing out the mistakes and providing constructive explanations. This will help students understand why an answer may be incorrect, reinforcing fundamental software engineering, data structure, and algorithm principles.

Prompt engineering will ensure that the ChatGPT API consistently provides accurate, contextualized feedback tailored to each student's unique needs and level of understanding (Hou et al., 2023).

**3.3 Assessments Evaluation with ChatGPT Prompt**: After collecting the students' assessment data and optimizing the ChatGPT prompt, we aim to leverage the power of ChatGPT to provide relevant, immediate, and constructive feedback to students for improved learning. For this purpose, the students' assessments will be evaluated using predefined rubrics to provide constructive feedback and mark the assessment based on its quality and correctness. The ChatGPT prompt will be provided with the student's assessments (the task asked and the student's solution) and evaluation criteria as input. Based on the supplied rubrics, the ChatGPT will return constructive feedback for the assessment and its overall score. The feedback will demonstrate the limitations and possible improvements for the student assessments. This will help module leaders of larger cohorts provide constructive, personalized feedback to the students by utilizing the flood of related expert knowledge of ChatGPT.

**3.4 Evaluation and Negotiation Process**: Although in the previous step of the proposed LLM-MATE approach, we have detailed prompt engineering and evaluation processes with ChatGPT for relevant, immediate, and constructive student feedback. However, AI-based systems, particularly LLM-based approaches, can generate useful content that appears authoritative and might be entirely or partially fabricated or unrelated to the query (Alier et al., 2024)(Khan et al., 2024). Such "hallucination" can mislead the students on feedback and might cause frustration and distrust. Therefore, we are interested in conducting a negotiation and validation experiment for the proposed LLM-MATE approach between the LLM and human experts. For this purpose, the feedback generated by the LLMs will be cross-validated by human experts (module leaders). The ChatGPT-generated results will be compared with the human-generated results for further improvements, and the evaluation rubrics will determine if there are any issues while generating the feedback. Additionally, the LLM-MATE approach will be explored for possible automation using ChatGPT API (Khan et al., 2024) to automate the learner's assessment feedback generation upon successful prompt-based feedback generation.

4. **Result and Discussion**

**Data selection**: To evaluate the performances of the proposed LLM-MATE approach in assessing students' submissions, we obtained 23 students consent out of 290 to use their submissions for ChatGPT-based analysis and generate personalize feedback. We have got comparatively low consent from the students. However, several requests have been communicated with the students using the learning management system (Canvas). One possible reason can be that we did not get a chance to communicate the importance of data collection to students in face-to-face teaching. Moreover, we obtained enough student consent to demonstrate the efficacy of the proposed LLM-MATE approach. We selected several students' assessments based on i) diverse sample size, i.e., highest, middle, and lowest obtained markings for the students who gave their consent, and ii) assessment sample marked by each module team member (the module involves four members). A software project case study was given in the assessment brief, and students were asked to develop a use case, class, and 3-tier architecture diagram, each weighting 30%, 30%, and 40%, respectively.

**ChatGPT prompting:** To generate more generalized results with the LLM-MATE approach, we experimented with ChatGPT, which involves updating certain parameters to fit a particular task, i.e., student assessment and a given dataset. For this purpose, various prompt engineering approaches exist in the literature, such as zero-shot, few-shot, chain-of-thought learning, etc. (Liu et al., 2023). Zero-shot learning generates outputs with LLMs with no examples provided as input along with the required task. In contrast, few-shot learning adds examples of the required task to the prompt, along with a task description and input (Chen et al., 2023). Similarly, chain-of-thought learning adds a chain of reasoning with the input task to

ChatGPT (Wei et al., 2022). We used zero-shot learning for the proposed LLM-MATE approach because of its widespread applications in literature, which require less input to generate desired outputs (Khan et al., 2024).

**Assessments Evaluation:** To evaluate the student's assessments with ChatGPT, we supplied the student assessment brief, pre-defined evaluation rubrics, and the student solution as input using a zero-shot prompt engineering approach. Also, an output format description is defined to evaluate the student assessment using pre-defined rubrics, generate constructive feedback comprising strengths and shortcomings that can be improved, and score the assessment out of 30 with justification. A sample used to interact with ChatGPT using zero-shot learning is shown in Figure 2. A 30 score is selected because we evaluate the use case diagram, weighting 30%. The prompt is designed by considering the proposed methodology elaborated in section 3.2. Similar prompts can be designed to evaluate the class and 3-tier architectural diagrams by replacing the evaluation rubrics and student submission solutions. In Figure 2, a short assessment brief is depicted, while for evaluation, a complete assessment brief was provided.

| Constructive and personalized feedback |
|---|
| Generate Constructive and personalized feedback by evaluating the student submission for the following assessment brief using pre-define rubrics in the following format |
| Strength of the assessment: |
| Shortcomings that can be improved: |
| Overall Score out of 30 and justification: |
| Assessment brief: A depot receives parcels for collection by customers. Designated staff record them in a sorted order in the depot system. The sorting method is predominantly by customer surname. Customers arrive at the depot and use a queuing system to collect their parcels. Once a customer collects their parcel, it is placed in a collected list and removed from those waiting to be collected. A parcel can be in two states, waiting for collection or collected........ |
| Pre-defined rubrics: The use case diagram uses standard notation - subject(system) boundary, actors and use cases. The elements are suitably derived from the scenario. There is use of <<extends>> & <<include>> relationship and the involved use cases are logical for this association |
| Student submission: Attached use case diagram image as student solution |

Figure 2: Zero-shot learning prompt to generate constructive feedback using the LLM-MATE approach.

The experiments to generate constructive feedback were conducted with GPT-4o, a model developed by OpenAI that analyzes text and images for insightful results. We use the web interface to access the GPT-4o model to evaluate the student assessments and generate constructive feedback using the ChatGPT website (ChatGPT4o, 2024). To evaluate the performance of the proposed LLM-MATE approach, we used the ground truth from the module team, who evaluated the student assessments manually for the Software Architecture submissions. In total, 4 teachers evaluated the student assessments, and the student consent obtained for this study covers assessment across all four module members. Therefore, we are also interested in identifying whether the LLM-MATE approach can help overcome the inconsistencies between the module team members when evaluating the student assessments. In the literature, it is considered one of the key challenges in providing uniform and consistent feedback to the students (Messer et al., 2024).

**RQ1: Performance of ChatGPT in Assessment Evaluation:** RQ1 aims to explore ChatGPT's performance in assisting the module team in generating personalized and constructive feedback by harnessing the power of LLM. Using the zero-shot learning prompt, as shown in Figure 2, to evaluate a learner assessment on software architecture and generate personalized and constructive feedback, the ChatGPT performs comparatively well in generating feedback by elaborating its strengths, shortcomings, overall score, and justification for the score. The feedback generated by ChatGPT is shown in Figure 3 based on a zero-shot learning prompt. The student assessment provided to ChatGPT for evaluation is completely anonymized and only provides the potential solution. It is evident from the feedback that ChatGPT can be a good alternative source for generating in-time, consistent, and personalized feedback when compared to the teaching team member-generated feedback, as shown in Figure 4, which summarizes the assessment without providing explicit details, i.e., strengths, limitation with possible improvements and rationale for marking. The ChatGPT-generated feedback is more personalized, explaining the strength of assessments, shortcomings that can improved, and justification for the evaluation and marks. The finding, therefore, supports the proposed research question and claim. It has been evident in the literature that AI can be useful for module leaders in generating feedback for a large cohort, and student learning improves by providing immediate feedback (González-Calatayud et al., 2021).

| Feedback Criteria | Detailed explanation |
|---|---|
| Strength of the Assessment | 1. **Standard Notation**: The student has used standard UML notation, including the system boundary, actors (Manager, Worker, Customer), and use cases. This demonstrates a good understanding of UML fundamentals.<br>2. **Logical Actor-Use Case Associations**<br>3. The relationships between actors and use cases are logical and align with the given scenario<br>   a. The **Manager** generates reports.<br>   b. The **Worker** manages parcels, customers, and updates parcel status.<br>   c. The **Customer** collects parcels.<br>4. **Use of Relationships**:<br>  • The diagram appropriately uses <<include>> and <<extend>> relationships:<br>    o <<include>> for shared functionality like "Remove Parcel" and "Remove Customer."<br>    o <<extend>> for optional functionality like "Apply Discount." |
| Shortcomings that Can Be Improved | 1. **Missing or Implicit Use Cases**:<br>  • **"Initialise List of Parcels and Customers"**: The scenario mentions initializing data from text files, but this use case is missing. It is a key feature of the system and should be included.<br>  • **"Find Parcel by ID"**: While parcel updates and removals are shown, a dedicated use case for searching parcels is not represented.<br>2. **Ambiguity in Some Relationships**:<br>  • The <<include>> relationships could be clarified further. For example:<br>    a. It is unclear why "Update Parcel Status" includes "Calculate Fee" but not "Collect Parcel" since collection logically triggers the fee calculation.<br>3. **Customer Actor Involvement**:<br>  • While the **Customer** actor is connected to "Collect Parcel," the use case could also indicate how the system interacts with customer data, such as maintaining a queue or adding/removing customers.<br>4. **Lack of Supporting Labels**:<br>  • Consider adding labels or notes to clarify decisions like why certain relationships are extended or included. This would improve readability. |
| Overall Score | 22/30 |
| Justification | The use case diagram demonstrates a solid understanding of UML standards, actors, and relationships. Logical associations are present, and the use of <<include>> and <<extend>> is appropriate in many areas. However:<br>• Key use cases such as "Initialize Data" and "Find Parcel by ID" are missing.<br>• Some relationships and actor interactions require further clarification or refinement.<br>• To improve, ensure all major functional requirements from the scenario are covered and relationships are clearly justified with supporting notes. |

Figure 3: Constructive feedback generated by ChatGPT using a zero-shot learning approach.

> You have done a good job with the UCD. The layout and notation are correct. I would suggest removing the Manager, as there are only two actors. The key missing functionality is loading customer data and loading parcel data from a file.

Figure 4: Feedback generated by the module team member.

**RQ2: Reliability of ChatGPT generated feedback:** To evaluate the reliability of the assessment feedback generated by ChatGPT, we conducted a validation study with the module team members involved in marking the same student assessments. For this purpose, we asked the module team to provide a confidence score ranging from 1 to 5 on the feedback generated by the ChatGPT, where 1-2 are referred to as low confidence, 3 are medium, and 4-5 are referred to as high confidence. Four teaching members were involved in the manual student assessment evaluation, and we conducted the validation study with the same module team. The team members were contacted by email to record their confidence in the ChatGPT-based assessment compared to their corresponding manual assessment. In the email, the teaching team was provided with the information on the student assessment feedback generated by the corresponding member and feedback generated by the ChatGPT. The participants were asked to provide a confidence score and any additional comments if there will be any. Moreover, a link to the original evaluation was provided in case the module team wanted to revise the student assessment. Interestingly, all four teaching members have shown high confidence in ChatGPT-based constructive and personalized feedback generation by providing a confidence value of 4, 5, 4, and 3, respectively. The average confidence score in ChatGPT-based feedback generation is 4. The findings suggest that ChatGPT can be a good alternative source of constructive, detailed, and personalized feedback generation. Additionally, it can improve student learning by providing detailed and constructive feedback on time (González-Calatayud et al., 2021). However, it still needs to be investigated by conducting experiments with the students.

To summarize, based on the results, ChatGPT can be a good alternative source for feedback generation. However, certain challenges when employing the LLM-MATE for feedback generation need further research and experiments. For example, ChatGPT evaluates assessments based on pre-defined rubrics. Therefore, it is required to develop evaluation rubrics carefully to cover all the possible aspects of the assessment brief. Also, ChatGPT is prone to hallucination and can generate feedback that is not relevant to assessments. Therefore, human experts should be added to the loop to overcome the hallucination factor (Alier et al., 2024)(Khan et al., 2024). Additionally, the results can be improved further by experimenting with a chain-of-thoughts prompting approach, which allows reasoning with the ChatGPT-generated results.

**Limitations:** With the proposed LLM-MATE approach, we obtained encouraging results demonstrating its potential to be used as an alternative source for generating constructive, detailed, and personalized feedback. However, the proposed approach does have several limitations that will be explored in the future, considering the limited word count. For example, we evaluated only a use case diagram to generate constructive feedback and evaluate with the domain experts. To better generalize the results, we aim to evaluate the class and 3-tier architecture diagram with the proposed approach and validate it from the domain experts. Also, the data collected is limited despite several requests from the students. Therefore, we aim to approach students during face-to-face teaching and ask for their consent to evaluate the proposed approach with a larger data set.

## 5. Conclusion and Future Work

In the proposed approach, we evaluate ChatGPT's performance in assessing student submissions and providing constructive feedback. The results show that ChatGPT can be a good alternative source for evaluating student submissions that could benefit module leaders of large cohorts and generate consistent feedback, unlike manual evaluators who use different templates to evaluate student assessments. In the literature, it has been emphasized that, in time, feedback can improve student learning. Therefore, the proposed approach can potentially improve student learning by providing timely, constructive, and personalized feedback.

In the future, the proposed approach can be extended to investigate how it can improve student learning by providing timely feedback. Also, few-shot and chain-of-though learning approaches will be investigated to improve its performance further. Moreover, the initial findings show great potential to automate the process to help module leaders provide immediate student feedback. Additionally, the proposed approach's performance will be further validated by exposing more software architecture diagrams.

**Ethical approval:** The study is conducted, and data are collected using the protocol number 0389 2024 Nov HSET.